\documentclass[runningheads]{llncs}
\usepackage{amsmath}
\usepackage{amssymb}
\usepackage{bm}        
\usepackage{multirow}
\usepackage{booktabs}    
\usepackage{multirow}    
\usepackage{makecell}    
\usepackage{xcolor}      
\usepackage{colortbl}    
\usepackage{marvosym}
\usepackage{graphicx,verbatim}
\definecolor{rowgray}{gray}{0.93}

\usepackage[T1]{fontenc}
\usepackage{graphicx,verbatim}
\begin{document}
\title{SC-TauPath: How Structural Connectivity Differences Shape Tau Distribution in Alzheimer's Disease}
\titlerunning{SC-TauPath}
%

\author{Jing Zhang\inst{1} \and
Norman Scheel\inst{2} \and
Minheng Chen\inst{1} \and
Tong Chen\inst{1} \and
Yanjun Lyu\inst{1} \and
David C. Zhu\inst{2} \and
Rong Zhang\inst{3} \and
Dajiang Zhu\inst{1}\textsuperscript{(\Letter)}}

\authorrunning{J. Zhang et al.}

\institute{University of Texas at Arlington, Arlington, TX, USA \\
\email{\{jxz7537,mxc2442,txc5603,yxl9168\}@mavs.uta.edu, dajiang.zhu@uta.edu} \and
Michigan State University, East Lansing, MI, USA \\
\email{scheelno@msu.edu, david.zhu@einsteinmed.edu} \and
University of Texas Southwestern Medical Center, Dallas, TX, USA \\
\email{rongzhang@texashealth.org}}
  
\maketitle             
\begin{abstract}
Tau neurofibrillary pathology in Alzheimer's disease (AD) follows stereotypical spatial patterns that are associated with white-matter connectivity. However, human brain structure varies substantially across individuals, thus understanding how these inter-individual differences in diffusion MRI(dMRI)-derived structural connectivity (SC) account for the corresponding heterogeneity in tau distribution remains unclear. We introduce \textbf{SC-TauPath}, a pairwise attribution
framework that maps SC differences directly to
tau pattern heterogeneity across subjects. For each pair of participants, SC-TauPath learns a mapping from pairwise SC differences
($\Delta$SC) to pairwise tau PET differences ($\Delta\tau$). The learned operator is then analytically decomposed to provide multi-scale attribution, identifying influential connections (edges), coordinated connection patterns (pathways), and key regions (hubs) most strongly associated with tau heterogeneity. Applied to 234 ADNI participants with dMRI-derived SC and $^{18}$F-Flortaucipir PET, SC-TauPath substantially outperforms the baseline by achieving higher cross-validated Pearson correlation in predicting inter-subject tau pattern differences. More importantly, our SC-TauPath captures SC-tau relationships not only across diagnostic groups, thereby revealing connectivity patterns associated with disease progression, but also within the same diagnostic stage, demonstrating that individual differences in SC contribute to tau heterogeneity even among subjects at similar clinical impairment levels. Code at \url{https://anonymous.4open.science/r/SC-TauPath-3482}



\end{abstract}
\section{Introduction}

Alzheimer’s disease (AD) is an irreversible and progressive neurodegenerative disorder and the leading cause of dementia worldwide. With no definitive cure currently existing, understanding the underlying mechanisms of disease progression remains critical for enabling timely intervention and improving therapeutic strategies. The pathological hallmarks of AD, intracellular tau neurofibrillary tangles are strongly associated with cognitive decline and disease progression~\cite{hyman2012national,de2025connectivity}. A defining feature of tau neurofibrillary pathology is the hierarchical spatial progression, first described by Braak and Braak~\cite{braak1991neuropathological}. Specifically, tau pathology starts in the entorhinal cortex (Braak stages~I-II), advances into limbic regions including the hippocampus (stages~III-IV), and is eventually widespread across the neocortex (stages~V-VI). Increasing evidence from in vivo brain imaging indicates that this progression is not random: tau accumulation follows patterns consistent with transsynaptic spread along white-matter connections~\cite{jacobs2018structural,strain2018loss,vogel2020spread}.

Motivated by this “prion-like” hypothesis of transsynaptic propagation, a range of computational models have been developed to characterize the relationship between diffusion tensor imaging diffusion MRI(dMRI)-derived structural connectivity (SC), which reflects white-matter architecture, and the spatial progression of tau pathology. The first example is the network diffusion model (NDM)~\cite{raj2012network}, which simulates tau spread over time on SC using a graph diffusion equation. Since then, numerous extensions and variants of the NDM have been proposed to examine how brain structural~\cite{fornari2019prion,fornari2020spatially,raj2021combined,yang2019longitudinal}, functional~\cite{brown2019patient,zhou2012predicting}, and multimodal networks~\cite{thompson2024combining,schoonhoven2023tau} relate to tau accumulation and disease progression. However, such approaches predominantly model how tau evolves over time in individual brain regions given a fixed SC.


In this work, we take a different perspective and propose \textbf{SC-TauPath}. Rather than simulating tau evolves, we aim to learn how differences in SC give rise to differences in tau deposition patterns across individuals. Specifically, our main contributions are: \textbf{(i) Novel pairwise-difference modeling.} To our knowledge, SC-TauPath is the first work that explicitly models inter-subject differences in SC as predictors of tau pathology differences, capturing variability both across and within diagnostic stages. \textbf{(ii) Multi-scale interpretability.} Through analytical decomposition of the learned operator, SC-TauPath provides interpretable attribution across influential connections (edges), coordinated connection patterns (pathways), and key regions (hubs) most strongly associated with tau heterogeneity. \textbf{(iii) Strong predictive performance.} We evaluate SC-TauPath on 234 ADNI participants. Compared with the classical NDM baseline, our method achieves a $3.1 \times$ improvement in predictive performance.


\begin{figure}[h!]
\centering
\includegraphics[width=0.90\columnwidth]{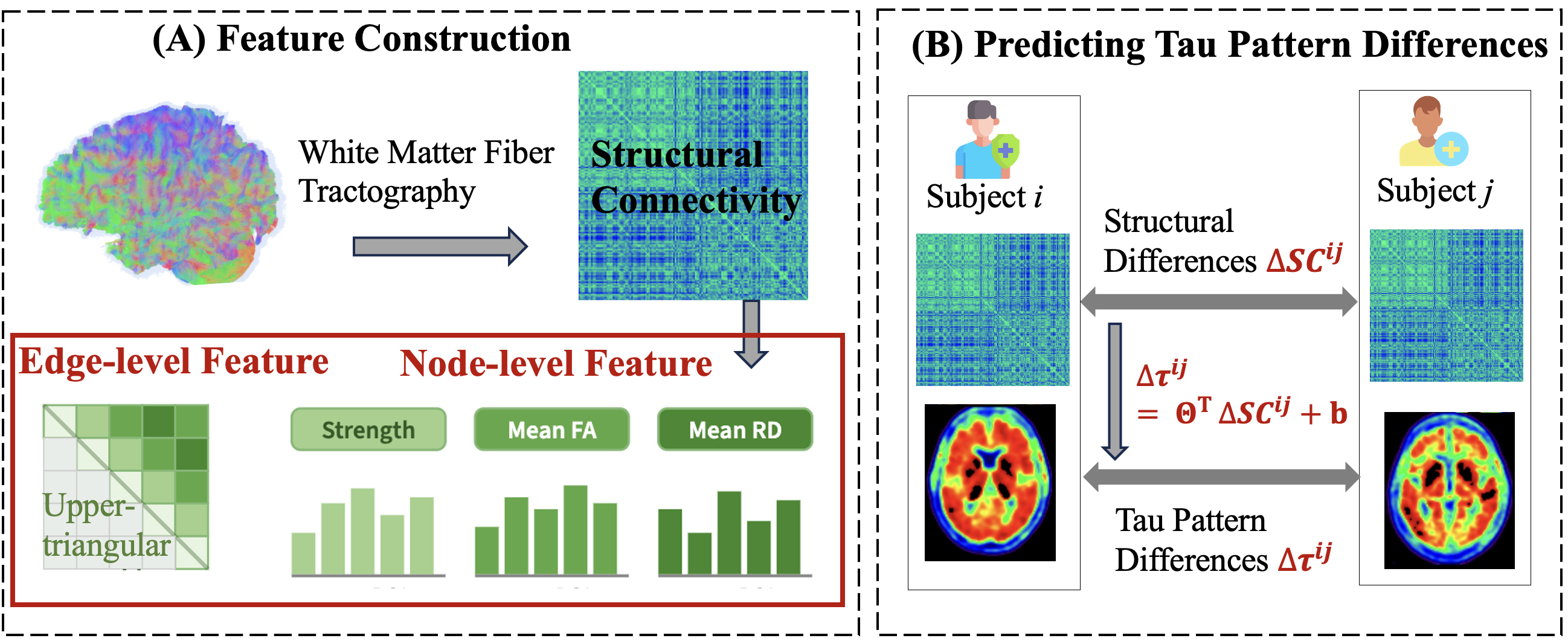}
\setcounter{figure}{0}
\caption{Overview of the proposed SC-TauPath with two main parts.}
\label{fig_main}
\end{figure}

\section{Methodology}
\subsection{Participants and Data Processing}
We analyzed 234 participants (122 CN, 83 MCI, and 29 AD) from the Alzheimer's Disease Neuroimaging Initiative (ADNI), each with tau PET ($^{18}$F-Flortaucipir), dMRI, and structural MRI (T1-weighted) scans available. Regional tau PET standardized uptake value ratios (SUVRs) were obtained following the standard pipeline in~\cite{villemagne2023centaur}. The PET image was first co-registered to the subject's T1 by SPM, subsequently, the T1 was normalized to the rrAD420 template - an anatomical atlas optimized for older adult populations - and the resulting transformation was applied to the PET image~\cite{scheel2022introducing,scheel2025functional}. SUVRs were computed voxel-wise using cerebellar gray matter as the reference region. Regional tau PET SUVRs were then extracted using the Brainnetome atlas(BNA)~\cite{fan2016human}, yielding a 246-dimensional SUVR vector per subject.

SC was derived following the standardized preprocessing pipeline described in~\cite{zhang2022predicting,zhang2021deep}. Briefly, skull stripping was applied to both normalized T1 and dMRI images, after which dMRI was registered to T1 space using FSL. The brain was parcellated into 246 regions of interest (ROIs) using the Brainnetome atlas, a fine-grained whole-brain parcellation grounded in structural connectivity architecture. Tractography yielded a subject-specific 246×246 connectivity matrix \(X\), whrere each element \(X_{ij}\) represents the edge-level connectivity weight between regions \(i\) and \(j\), quantified by fiber count. Regional node-level features were additionally extracted, including node strength, computed by $s_i = \sum_{j} X_{ij}$, mean fractional anisotropy (FA), and mean radial diffusivity (RD) (See (Fig~\ref{fig_main}(A)). All features were z-score normalized prior to model input. 

\subsection{Pairwise Regression Framework}
\label{sec:pairwise}


\textbf{Pairwise Difference Formulation.} Let $\mathcal{S}=\{1,\dots,N\}$ denote the subject set. For each subject
$s\in\mathcal{S}$, we extract (i)~a SC feature
vector $\mathbf{x}_s\in\mathbb{R}^{d}$ and (ii)~a regional tau PET SUVR
vector $\boldsymbol{\tau}_s\in\mathbb{R}^{R}$, where $R=246$ and
$d=30{,}873$ comprises $30{,}135$ upper-triangular edge-level features and $738$ node-level features. For any ordered pair $(i,j)$ with $i\neq j$, we
define$  \Delta \mathbf{x}_{ij} = \mathbf{x}_i - \mathbf{x}_j,
  \Delta \boldsymbol{\tau}_{ij} = \boldsymbol{\tau}_i - \boldsymbol{\tau}_j$,
%
%
resulting the antisymmetry
$\Delta \mathbf{x}_{ji}=-\Delta \mathbf{x}_{ij}$ and
$\Delta \boldsymbol{\tau}_{ji}=-\Delta \boldsymbol{\tau}_{ij}$. During
training, both orderings $(i,j)$ and $(j,i)$ are included as data
augmentation, while only upper-triangular pairs are retained at evaluation. Each pair is assigned a diagnostic type based on the
ordered labels $(\ell_i,\ell_j)$ with
$\ell \in \{\mathrm{CN},\mathrm{MCI},\mathrm{AD}\}$, yielding six pair categories: CN-CN, CN-MCI, CN-AD, MCI-MCI, MCI-AD, and AD-AD.


\noindent \textbf{Regularized Linear Mapping.} The SC feature space is high-dimensional ($d\gg N$), we
therefore apply principal component analysis (PCA) that project each subject into a reduced subspace: $  \mathbf{z}_s
  = \mathbf{P}^{\top}(\mathbf{x}_s-\boldsymbol{\mu}),
  \mathbf{P}\in\mathbb{R}^{d\times p}$,
%
%
where $\boldsymbol{\mu}$ is the training-set mean and
$p=\mathrm{rank}(\mathbf{X}_{\mathrm{train}})$. Pairwise inputs are then $  \Delta \mathbf{z}_{ij}
  = \mathbf{P}^{\top}(\mathbf{x}_i - \mathbf{x}_j)
  \in \mathbb{R}^{p}$.
%
%
We train a regularized linear model to predict tau differences:
\begin{equation}
  \Delta \boldsymbol{\hat{\tau}}_{ij}
  =
  \mathbf{W}\,\Delta \mathbf{z}_{ij} + \mathbf{b},
  \qquad
  \mathbf{W}\in\mathbb{R}^{R\times p},\;
  \mathbf{b}\in\mathbb{R}^{R}
  \label{eq:lin_map}
\end{equation}
Parameters $(\mathbf{W},\mathbf{b})$ are estimated by Ridge regression:
\begin{equation}
  \min_{\mathbf{W},\mathbf{b}}
  \sum_{(i,j)}
  \left\|
    \Delta \boldsymbol{\tau}_{ij}
    - \mathbf{W}\Delta \mathbf{z}_{ij}
    - \mathbf{b}
  \right\|_2^2
  + \alpha \|\mathbf{W}\|_F^2
  \label{eq:ridge_obj}
\end{equation}
where $\alpha$ is selected via validation-set mean per-pair Pearson correlation:
\begin{equation}
  \mathrm{score}(\alpha)
  =
  \frac{1}{|\mathcal{P}_{\mathrm{val}}|}
  \sum_{(i,j)\in\mathcal{P}_{\mathrm{val}}}
  \mathrm{corr}\!\left(
    \Delta \boldsymbol{\hat{\tau}}_{ij},
    \Delta \boldsymbol{\tau}_{ij}
  \right)
  \label{eq:val_score}
\end{equation}

\subsection{Attribution Analysis}
\label{sec:attr}


\textbf{Effective Linear Operator. }Because the proposed framework defines a fully linear mapping from SC
differences to tau differences, attribution can be derived analytically
by composing the PCA projection and Ridge coefficients into a single
operator in the original feature space. Recall that PCA projects each
subject's SC feature vector as
$\mathbf{z}_s = \mathbf{P}^{\top}(\mathbf{x}_s - \boldsymbol{\mu})$,
and Ridge regression estimates a weight matrix $\mathbf{W}$ such that
$\Delta\boldsymbol{\hat{\tau}}_{ij} = \mathbf{W}\,\Delta\mathbf{z}_{ij}$.
Substituting the PCA projection yields:
\begin{equation}
  \Delta \boldsymbol{\hat{\tau}}_{ij}
  =
  \mathbf{W}\,\mathbf{P}^{\top}\,\Delta \mathbf{x}_{ij}
  \label{eq:compose_operator}
\end{equation}
We define the effective operator in the original SC feature space as:
\begin{equation}
  \boldsymbol{\Theta}
  \;\triangleq\;
  \mathbf{W}\,\mathbf{P}^{\top}
  \;\in\; \mathbb{R}^{R \times d}
  \label{eq:theta_operator}
\end{equation}
so that $\Delta\boldsymbol{\hat{\tau}}_{ij} =
\boldsymbol{\Theta}\,\Delta\mathbf{x}_{ij}$. Each entry $\Theta_{r,k}$
quantifies the linear contribution of SC feature~$k$ to the predicted
tau difference at ROI~$r$.


\noindent \textbf{Edge-Level Attribution.} To obtain a scalar importance score for each SC edge, we aggregate its influence across all tau target regions. For edge~$k$:
\begin{equation}
  \mathrm{Imp}(k)
  =
  \sum_{r=1}^{R}
  \bigl|\Theta_{r,k}\bigr|
  =
  \bigl\|\boldsymbol{\Theta}_{\cdot,k}\bigr\|_{1}
  \label{eq:edge_importance}
\end{equation}
This $\ell_1$ aggregation captures the total magnitude of an edge's
influence on inter-subject tau differences. Edges are ranked by $\mathrm{Imp}(k)$ to identify the dominant
connectivity pathways associated with tau variability.


\noindent\textbf{Hub-Level Attribution.} To identify brain regions that serve as convergence points of important
connections, we define a hub score for each ROI~$r$ as the cumulative
importance of all incident edges. Let $\mathcal{E}(r)$ denote the set of
edges incident to ROI~$r$:
\begin{equation}
  \mathrm{Hub}(r)
  =
  \sum_{k \in \mathcal{E}(r)} \mathrm{Imp}(k),
  \qquad
  \widetilde{\mathrm{Hub}}(r)
  =
  \frac{\mathrm{Hub}(r)}
       {\max_{r'}\,\mathrm{Hub}(r')}
  \label{eq:hub_norm}
\end{equation}
High $\widetilde{\mathrm{Hub}}$ scores identify ROIs whose structural
connections collectively exert the strongest influence on inter-subject
tau differences.


\noindent \textbf{Stage-Dependent Operator Analysis.} To examine whether connectivity–tau relationships vary across disease
stages, we independently estimate the regression operator within each
diagnostic pair subset. Let $\mathcal{P}_c$ denote the set of training pairs belonging to
diagnostic category $c$.
Fitting the same pairwise regression model on $\mathcal{P}_c$ yields a
stage-specific operator $  \boldsymbol{\Theta}^{(c)}.$ Stage-dependent edge importance is then computed as

\begin{equation}
  \mathrm{Imp}^{(c)}(k)
  =
  \sum_{r=1}^{R}
  \left|
  \Theta^{(c)}_{r,k}
  \right|.
  \label{eq:stage_edge_importance}
\end{equation}

Comparing $\boldsymbol{\Theta}^{(c)}$ across $c$ reveals how structural
drivers of tau variability evolve along the Alzheimer’s disease spectrum.

\section{Experiments and Results}

\subsection{Experimental Setup}
For the model configuration, PCA was fitted on the fit partition of each fold, retaining the full
rank. The
Ridge regularization parameter $\alpha$ was selected from a
logarithmically spaced grid ($10^2$ to $10^9$) by maximizing the
validation-set mean per-pair Pearson $\bar{r}$
(Eq.~\ref{eq:val_score}), and the $\alpha = 10^7$ was
consistently selected across all five folds. We employed 5-fold subject-level cross-validation. At each fold, 20\%
of subjects were held out for testing, with the remaining 80\% split
into fit (85\%) and validation (15\%) partitions. Training pairs included both orderings $(i,j)$ and $(j,i)$ as data
augmentation; validation and test pairs used each unordered pair only
once. Test-set predictions from all five folds
were assembled into an out-of-fold (OOF) collection of held-out pairs. For each pair $(i,j)$,
the Pearson $r$ between the predicted
$\Delta\boldsymbol{\hat{\tau}}_{ij}$ and observed
$\Delta\boldsymbol{\tau}_{ij}$ was computed across all  ROIs; the mean over all OOF pairs, $\bar{r}$, served as the primary performance metric.

\subsection{Results}



\textbf{SC Topology Predicts Inter-Individual Tau Differences.} We first evaluated whether inter-individual SC differences $\Delta\mathbf{z}^{\mathrm{SC}}$ could predict inter-individual tau PET differences $\Delta\boldsymbol{\tau}$ across all diagnostic pair types,  the result summarized in Table~\ref{tab:ablation}. The first two rows show the results of our SC-TauPath and the baseline NDM, respectively. In our experimental, NDM formulates tau propagation as heat-equation diffusion on the subject-specific SC network, seeded at the bilateral entorhinal cortex, which represents the canonical Braak stage I/II origin of tau pathology~\cite{braak1991neuropathological}. For each subject, graph Laplacian eigenmodes were computed from the SC, and diffusion parameters, including the spreading rate and diffusion time were optimized via grid search on each training fold. The resulting per-subject NDM predictions
$\mathbf{x}^{\mathrm{NDM}} \in \mathbb{R}^{246}$
were subsequently used to construct pairwise difference features
$\Delta\mathbf{x}^{\mathrm{NDM}} = \mathbf{x}_i^{\mathrm{NDM}} -
\mathbf{x}_j^{\mathrm{NDM}}$ for downstream regression.

\begin{table*}[t]
\centering
\caption{Out-of-fold prediction performance (Pearson $r$, mean\,$\pm$\,std 
over 5 folds) across diagnostic pair types. \textbf{Bold} denotes the best 
value per column (ties shared). $^{*}$All predictions significantly exceed chance (permutation test, $n=2{,}000$ subject-level label permutations; $p < 0.001$ for all pair types.)}
\label{tab:ablation}
\resizebox{\textwidth}{!}{%
\begin{tabular}{ll ccccccc}
\toprule
\multirow{2}{*}{\textbf{Model}} &
\multirow{2}{*}{\textbf{Input}} &
\multicolumn{7}{c}{\textbf{Pearson $r$ (mean\,$\pm$\,std, 5-fold CV)}} \\
\cmidrule(lr){3-9}
& &
\makecell{\textbf{Overall} \\ ($n$\,=\,5359)} &
\makecell{\textbf{CN-CN} \\ ($n$\,=\,1468)} &
\makecell{\textbf{CN-MCI} \\ ($n$\,=\,1927)} &
\makecell{\textbf{CN-AD} \\ ($n$\,=\,719)} &
\makecell{\textbf{MCI-MCI} \\ ($n$\,=\,710)} &
\makecell{\textbf{MCI-AD} \\ ($n$\,=\,462)} &
\makecell{\textbf{AD-AD} \\ ($n$\,=\,73)} \\
\midrule
\rowcolor{rowgray}
SC-TauPath$^{*}$
  & $\Delta\mathbf{z}^{\mathrm{SC}}$
  & $\mathbf{0.236 \pm 0.040}$
  & $\mathbf{0.221 \pm 0.050}$
  & $\mathbf{0.241 \pm 0.028}$
  & $\mathbf{0.241 \pm 0.149}$
  & $\mathbf{0.188 \pm 0.041}$
  & $0.161 \pm 0.087$
  & $0.158 \pm 0.089$ \\
NDM$^{*}$
  & $\Delta\mathbf{x}^{\mathrm{NDM}}$
  & $0.077 \pm 0.042$
  & $0.069 \pm 0.103$
  & $0.053 \pm 0.057$
  & $0.172 \pm 0.106$
  & $0.001 \pm 0.079$
  & $0.115 \pm 0.064$
  & $0.091 \pm 0.082$ \\
SC-TauPath+NDM$^{*}$
  & $[\Delta\mathbf{z}^{\mathrm{SC}} \| \Delta\mathbf{x}^{\mathrm{NDM}}]$
  & $0.235 \pm 0.040$
  & $\mathbf{0.221 \pm 0.051}$
  & $0.240 \pm 0.028$
  & $\mathbf{0.241 \pm 0.150}$
  & $0.186 \pm 0.041$
  & $\mathbf{0.161 \pm 0.088}$
  & $\mathbf{0.161 \pm 0.092}$ \\
\bottomrule
\end{tabular}%
}
\end{table*}

\begin{figure}[t]
\centering

\begin{minipage}{0.50\textwidth}
\setcounter{figure}{1}
    \centering
    \includegraphics[width=\linewidth]{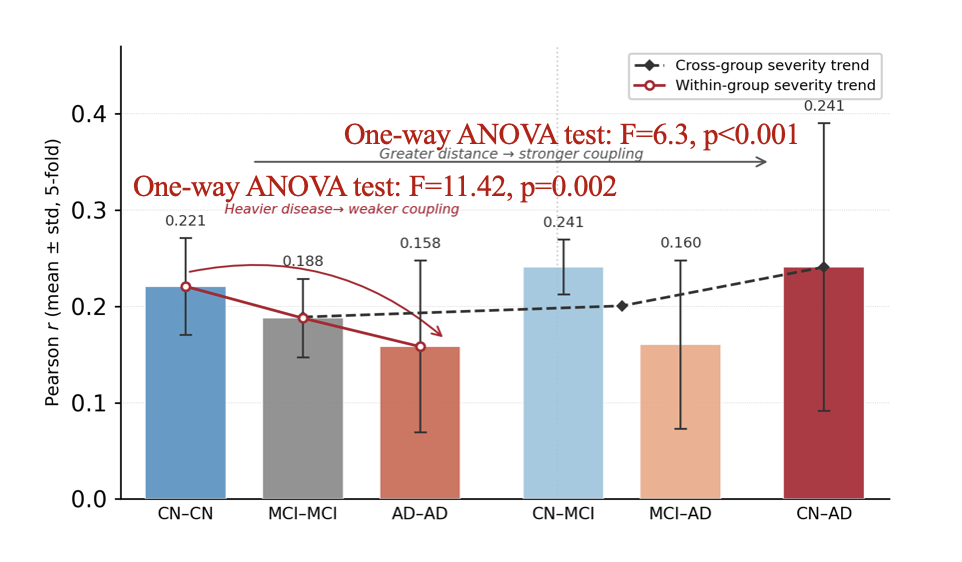}
    \caption{SC-tau Prediction performance, which progressively increases with cross-stage (the black upward dashed curve) and decreases within-stage from CN to AD (the red downward solid curve).}
    \label{fig:trend}
\end{minipage}
\hfill
\begin{minipage}{0.40\textwidth}
\setcounter{figure}{3}
    \centering
    \includegraphics[width=\linewidth]{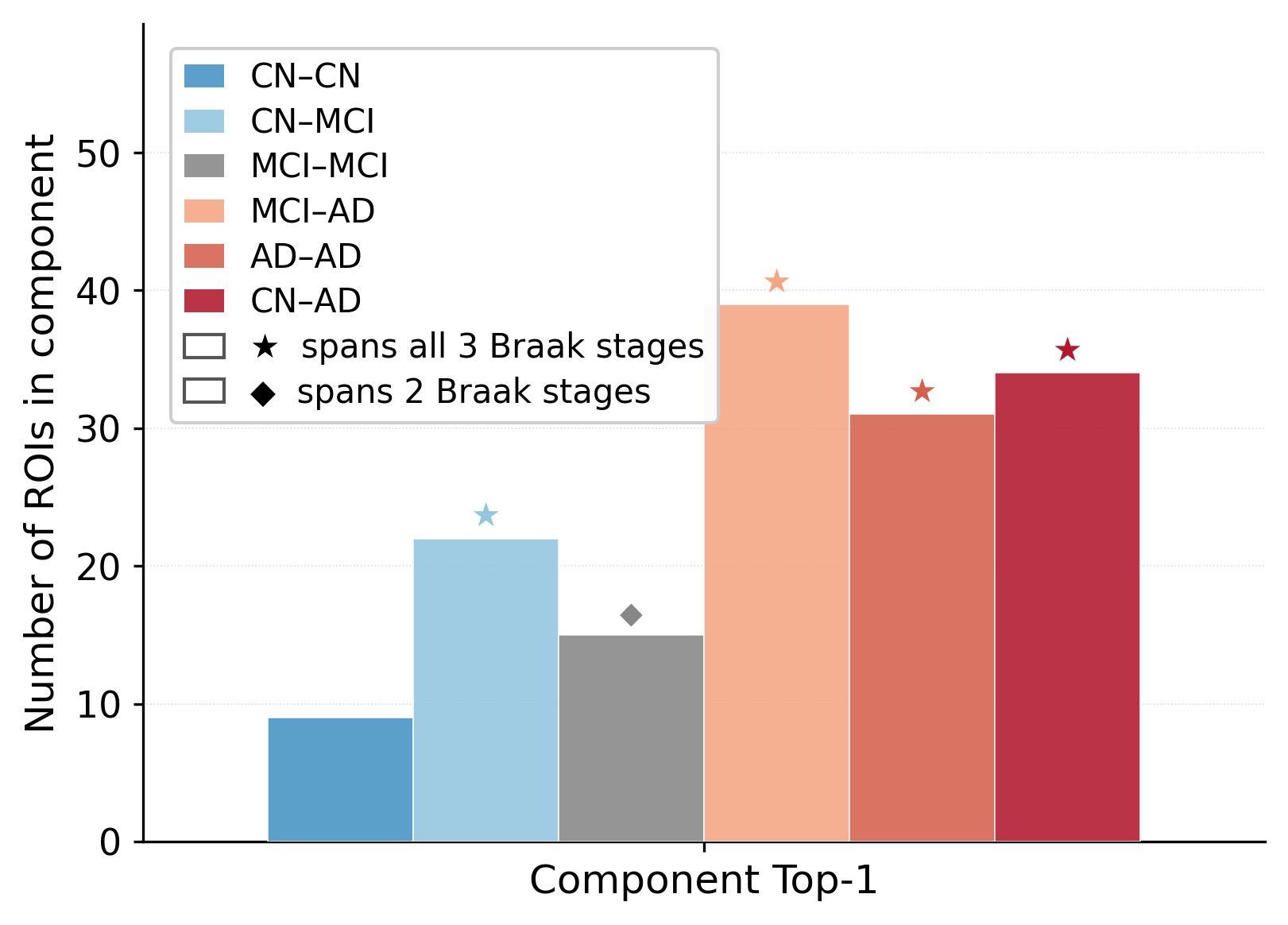}
    \caption{Connected components of top-50 attribution edges. The largest connected component is larger in AD-related pairs than in CN-CN pairs.}
    \label{fig:components}
\end{minipage}

\end{figure}

As shown in table~\ref{tab:ablation}, NDM yielded markedly lower overall performance ($r = 0.077 \pm 0.042$; 3.1$\times$ below our method), indicating that physics-based diffusion carry limited information about inter-individual tau variability. Here, Pearson correlation measures the spatial agreement between predicted and observed $\Delta\tau$ patterns, with higher values reflecting better predictive performance. In contrast, our method directly leveraging pairwise differences in SC topology, achieving an overall Pearson $r = 0.236 \pm 0.040$, confirming that inter-individual differences in SC carry predictive information for tau spatial heterogeneity.

We further evaluated the performance varied across diagnostic pair types. As expected, cross-stage contrasts such as CN-AD pairs yielded the highest correlations ($r = 0.241 \pm 0.149$), reflecting large between-group differences in both structural topology and tau burden. Importantly, however, within-group comparisons showed reliable SC-tau coupling: CN-CN ($r = 0.221 \pm 0.050$), MCI-MCI ($r = 0.188 \pm 0.041$), and AD-AD ($r = 0.158 \pm 0.089$) all exceeded NDM baseline, demonstrating that individual differences in SC topology account for inter-individual tau variability independent of disease stage. We further examine whether NDM provides complementary predictive information by combining NDM with our method. However, the result (last row in Table~\ref{tab:ablation}) showed a slight decrease in performance, suggesting that SC-TauPath already captures the dominant SC-constrained information of tau variability.


Moreover, we have two interesting findings, as show in Fig.~\ref{fig:trend}: first, prediction performance increased cross-group, (See black upward dashed curve in Fig.~\ref{fig:trend}). This trend indicates that SC differences more related with tau differences when paired individuals are separated by larger contrasts in disease-stage. Second, prediction performance progressively decreased across within-group, (CN-CN > MCI-MCI > AD-AD, See red downward solid curve in Fig.~\ref{fig:trend}). One possible interpretation is that tau pathology remains in an early or evolving phase, during which differences in SC can meaningfully shape tau patterns. In contrast, at later disease stages, tau pathology may reach a more spatially stable state, resulting in reduced variability. Both trends were statistically significant.



\noindent \textbf{Interpretation.} Beyond predictive performance, an effective model should provide interpretable insights into which structural connections drive inter-individual tau differences, thereby offering neurobiological insight into the SC-tau relationship. Using the attribution method described in Section~\ref{sec:attr}, we first quantified the importance of each SC edge and identified the top-ranked connections for CN-AD pairs. Building upon these edge-level importance scores, we further
characterized structural organization by analyzing attribution pathways and hub regions. The Top-5 CN-AD attribution edges, pathways, and high-importance hubs are illustrated in Fig.~\ref{fig_result}. For each edge and hub, importance scores are shown across CN-CN, CN-AD, and AD-AD pairs, enabling direct comparison of disease-stage specificity. There are several findings:

From the \textbf{\textit{Attribution Edge }}perspective (See Fig~\ref{fig_result} (a)), first, the inferior temporal gyrus (ITG[L], BNA area A20cv, Braak~III-IV) appears in three of the five top-ranked edges, connecting to the parahippocampal gyrus (PhG[R], edges~\#1 and~\#2) and the lateral occipital cortex (LOcC[R], edge~\#4). Additionally, the top two edges both connect ITG[L] to distinct
subregions of the right parahippocampal gyrus: the posterior parahippocampal cortex (PhG[R], area~TL, edge~\#1) and the medial parahippocampal cortex (PhG[R], area~TH, edge~\#2), respectively. This recurrent involvement suggests that the ITGis structurally associated with the preferential expression of CN-AD tau divergence, consistent with neuropathological staging that the inferior temporal cortex occupies an intermediate Braak III–IV position between medial temporal epicentres and widespread neocortical involvement~\cite{lowe2018widespread}. Moreover, across all five edges, importance scores for CN-AD pairs were $14$-$30\times$ higher than those of the same edges in CN-CN pairs (edge~\#1: $30.2\times$), indicates that these connections are selectively associated with disease-related tau accumulation.


\begin{figure}[h!]
\centering
\includegraphics[width=0.85\columnwidth]
{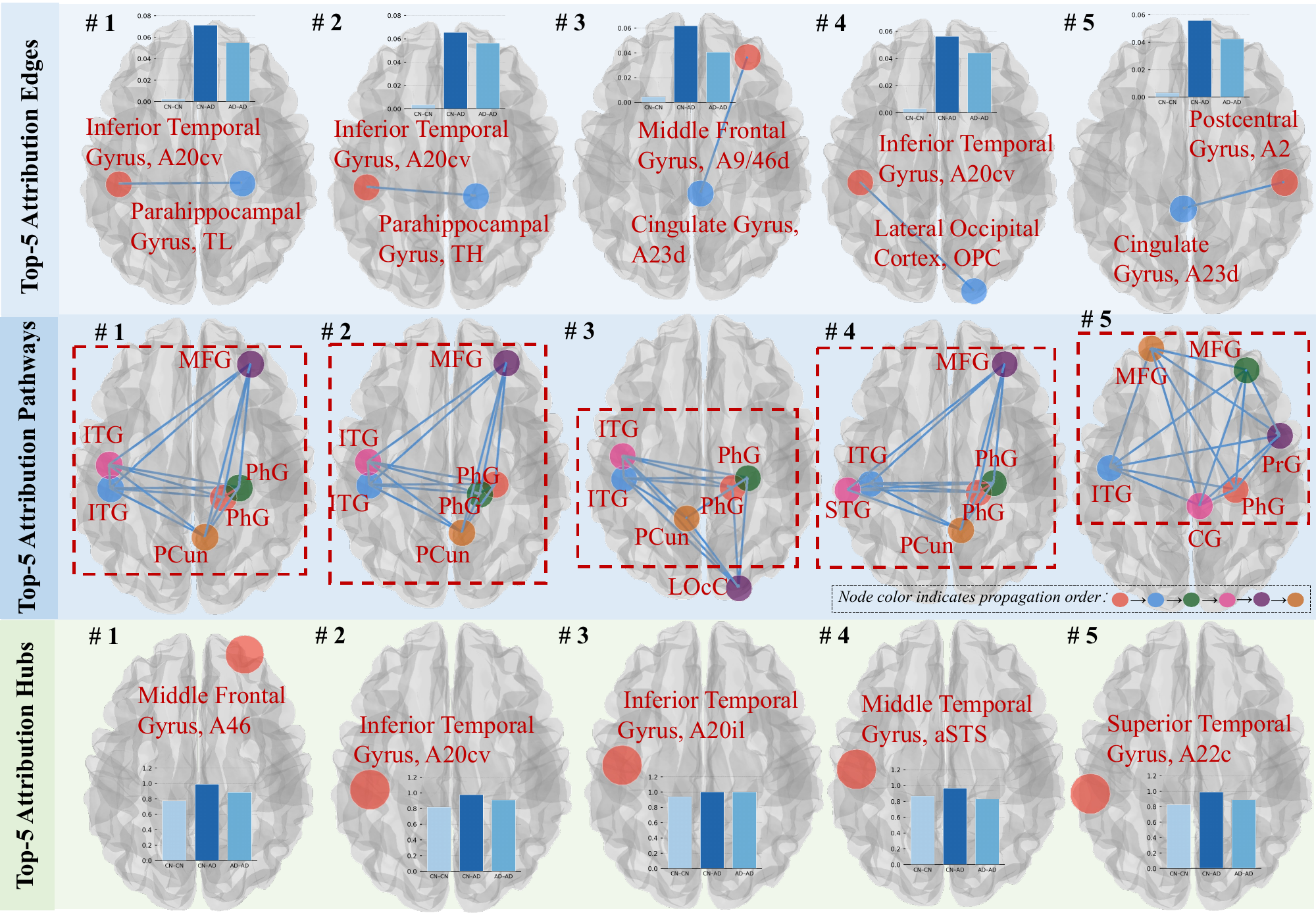}
\setcounter{figure}{2}
\caption{
Multi-scale structural determinants of CN-AD tau differences:
\textbf{(a)} top attribution edges,
\textbf{(b)} top attribution pathways,
and \textbf{(c)} top-importance hubs.
}
\label{fig_result}
\end{figure}


We further identified \textbf{\textit{multi-hop structural pathways}} that jointly contribute to inter-individual tau differences, recognizing that tau-related heterogeneity in the highly interconnected brain is more likely driven by coordinated network interactions than by isolated connections. Specifically, we constructed a subgraph from the top-200 attribution edges for CN-AD pairs and enumerated all simple paths (i.e., paths visiting each ROI at most once) from Braak~I-II source regions to Braak~V-VI target regions, ranking each path by its cumulative edge attribution score. The resulting top-5 attribution paths are visualized in Figure~\ref{fig_result}(b), where node color encodes propagation order. All five top-ranked pathways originate from the right parahippocampal gyrus (PhG[R], Braak~I-II) and converge on a shared structural core: the bidirectional PhG[R]$\leftrightarrow$ITG[L] axis, highlighted by the dashed red box in Fig.~\ref{fig_result}(b), suggesting that SC along the medial-to-lateral temporal axis is associated with CN-AD tau differences. We also analyzed connected components within the top-50 attribution-edge subgraph, As shown in Fig.~\ref{fig:components}. The largest component is markedly larger in AD-related pairs than in CN-CN pairs, whose edges are fragmented, indicating brain structural organization associated with disease-related tau differences.

Furthermore, we examined \textbf{\textit{attribution hubs}} derived from aggregated edge importance (Fig.~\ref{fig_result}(c)). The highest-ranked hub was the right middle frontal gyrus (MFG[R], normalized hub score = 1.000). Notably, MFG[R] was also consistently identified in the top-ranked attribution edges and pathways (Fig.~\ref{fig_result}(a-b)), providing convergent multi-scale evidence for its  association with inter-individual tau differences.


\section{Conclusion}

We introduced SC-TauPath, a pairwise attribution framework that maps inter-subject SC differences to tau distribution heterogeneity. By analytically decomposing a regularized
linear operator into edge-, pathway-, and hub-level scores, the framework produces multi-scale attribution maps that align well with Braak staging patterns. Beyond predictive performance, SC-TauPath provides neurobiologically grounded interpretability, linking specific SC to downstream tau divergence in both cross and within diagnostic groups. Future work will extend the framework to longitudinal data, graph neural network architectures, and multimodal inputs including functional connectivity.


    



%
%

\bibliographystyle{splncs04}
\bibliography{ref}

\end{document}